\documentclass[aps,prl,twocolumn,preprintnumbers,amsmath,amssymb,superscriptaddress]{revtex4-2}
\usepackage{graphicx}
\usepackage[caption=false]{subfig}
\usepackage{epsfig}
\usepackage{dcolumn}
\usepackage{bm}
\usepackage{color}
\usepackage{float}
\usepackage[colorlinks]{hyperref}
\usepackage{verbatim}
\usepackage{algorithm}
\usepackage{algorithmic}

\usepackage{booktabs}

\usepackage{multirow}

\hypersetup{citecolor = blue}

\def\bra#1{\left\langle#1\right|}
\def\ket#1{\left|#1\right\rangle}

\def\be{\begin{equation}}       \def\ee{\end{equation}}
\def\bea{\begin{eqnarray}}      \def\eea{\end{eqnarray}}
\def\ba{\begin{array}}
	\def\ea{\end{array}}
\def\bnum{\begin{enumerate} }
	\def\enum{\end{enumerate}}

\def\=>{\Rightarrow}
\def\>{\rightarrow}

\def\eye2{Fathbb{I}}

\def\Eq#1{Eq.~(\ref{#1})}

\renewcommand{\c}[1]{{\cal #1}}

\renewcommand{\>}{\rangle}

\newcommand{\eq}[2]{
	\begin{equation}
		#1 \label{#2}
	\end{equation}
}

\renewcommand{\rm}[1]{\mathrm{#1}}

\usepackage{listings}
\usepackage{color}
\definecolor{lightgray}{gray}{1}

\begin{document}
	
\title{Variational post-selection for ground states and thermal states simulation}

\author{Shi-Xin Zhang}
\thanks{The two authors contributed equally to this work.} 
\email{shixinzhang@tencent.com}
\affiliation{Tencent Quantum Laboratory, Tencent, Shenzhen, Guangdong 518057, China}

\author{Jiaqi Miao}
\thanks{The two authors contributed equally to this work.} 
\affiliation{School of Physics, Zhejiang University, Hangzhou, Zhejiang 310027, China}

\author{Chang-Yu Hsieh}
\email{kimhsieh@zju.edu.cn}
\affiliation{Innovation Institute for Artificial Intelligence in Medicine of Zhejiang University, College of Pharmaceutical Sciences, Zhejiang University, Hangzhou, 310058, China}

\begin{abstract}
    Variational quantum algorithms (VQAs), as one of the most promising routes in the noisy intermediate-scale quantum (NISQ) era, offer various potential applications while also confront severe challenges due to near-term quantum hardware restrictions. In this work, we propose a framework to enhance the expressiveness of variational quantum ansatz by incorporating variational post-selection techniques. These techniques apply variational modules and neural network post-processing on ancilla qubits, which are compatible with the current generation of quantum devices. Equipped with variational post-selection, we demonstrate that the accuracy of the variational ground state and thermal state preparation for both quantum spin and molecule systems is substantially improved. Notably, in the case of estimating the local properties of a thermalized quantum system, we present a scalable approach that outperforms previous methods through the combination of neural post-selection and a new optimization objective.
\end{abstract}

\date{\today}
\maketitle

\section{Introduction} 
    Variational quantum algorithms (VQAs) \cite{Cerezo2020b, Bharti2021, Endo2020, Lennart2021} have attracted great attention due to the moderate requirements on quantum hardware resources and are promising candidates towards practical quantum advantage \cite{Arute2019, Zhong2020} in the noisy intermediate-scale quantum (NISQ) era \cite{Preskill2018}. Specifically, ground state and thermal state preparation via VQAs are under extensive investigation, where the former is often denoted as the variational quantum eigensolver (VQE) \cite{Peruzzo2014, OMalley2016, McClean2016, Kandala2017, Liu2019b, McArdle2020, Grimsley2019, Hsieh2019, Liu2023a, Liu2021c} while the latter is often called Gibbs state preparation \cite{Wu2019a, Motta2020, Verdon2019, Liu2019, Chowdhury2020, Wang2020b, sewell2022, warren2022, chen2023, Rall2023, consiglio2023, Wang2023, consiglio2023variational, shtanko2023}. However, these VQAs are still very restricted in terms of expressive power due to the limitation of current-generation quantum hardware. Therefore, it is urgent and necessary to explore useful techniques that can enhance VQAs with the consideration of near-term quantum hardware.

    In this work, we introduce an important quantum algorithm component that we call variational post-selection to boost the potential of VQAs and chart a promising path to demonstrate quantum advantages on near-term quantum hardware. Post-selection has been extensively explored in the design of quantum algorithms, including the linear combination of unitaries \cite{Childs2012}, quantum signal processing \cite{Low2017}, and probability implementation of imaginary time evolution \cite{Liu2020, Lin2020}. Post-selection has also found relevance in quantum physics such as measurement-induced entanglement phase transitions and space-time dual circuits  \cite{Li2018b, Chan2019, Skinner2019a, Li2019b, Choi2020, Jian2020a, Tang2021, Lavasani2021a, Ippoliti2020, Ippoliti2020a, Lu2021, Ippoliti2021, Garratt2023, Liu2023c, Liu2024}. Since extensive post-selection in general requires exponential resources regarding the number of measurement qubits, we must be careful to introduce a post-selection scheme into the VQA context with acceptable overhead. Besides, inspired by previous works attempting to incorporate neural networks with parameterized quantum circuits \cite{Liu2018d, Liu2019, Verdon2019, Hsieh2019, Benedetti2021, Rivera-Dean2021, Torlai2020, Bennewitz2021, Zhang2021b, Zhang2021d, Zhang2021neural, miao2023}, we combine neural network post-processing with variational quantum computing from a new angle, utilizing the idea of post-selection in some VQA scenarios.
	
\section{Results} 
    \subsection{Setup}
\begin{figure}[t]\centering
    \includegraphics[width=0.48\textwidth]{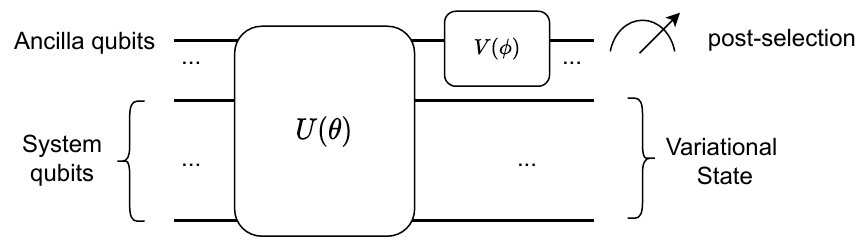}
    \caption{Schematic illustration on the VQE with our proposed variational post-selection scheme. The ancilla qubits go through some further variational circuit module $V(\phi)$ so that the effective post-selection output is also automatically adjusted.}
    \label{fig:vqe-post}
\end{figure}

    Our proposal constructs the variational quantum ansatz with several ancilla qubits and applies post-selection to these ancilla qubits. The basic scheme for variational post-selection in the ground state preparation problem is shown in Fig.~\ref{fig:vqe-post}. The scheme is termed `variational post-selection' due to the additional variational quantum module $V(\phi)$ applied to the ancilla qubit before conducting post-selection measurements. This module can be adjusted by tuning variational parameters $\phi$ such that the post-selection output is automatically optimized and determined. The post-selection overhead due to failed measurement outputs can be alleviated by limiting the number of ancilla qubits and customizing the objective functions. The success probability for the post-selection in the following numerical results indicates that the post-selection overhead is indeed scalable for practical VQE tasks. In sum, the variational post-selection scheme, as described, is to run the VQE with ancilla qubits, and only the results of ancilla qubits measurement outputs consistent with predefined outputs are retained and used to evaluate the expectation on system qubits. The expectation is further used as an objective function and the circuit parameters $\theta$ and $\phi$ are iteratively adjusted to minimize this objective.
	
\begin{figure}[t]\centering
    \includegraphics[width=0.47\textwidth]{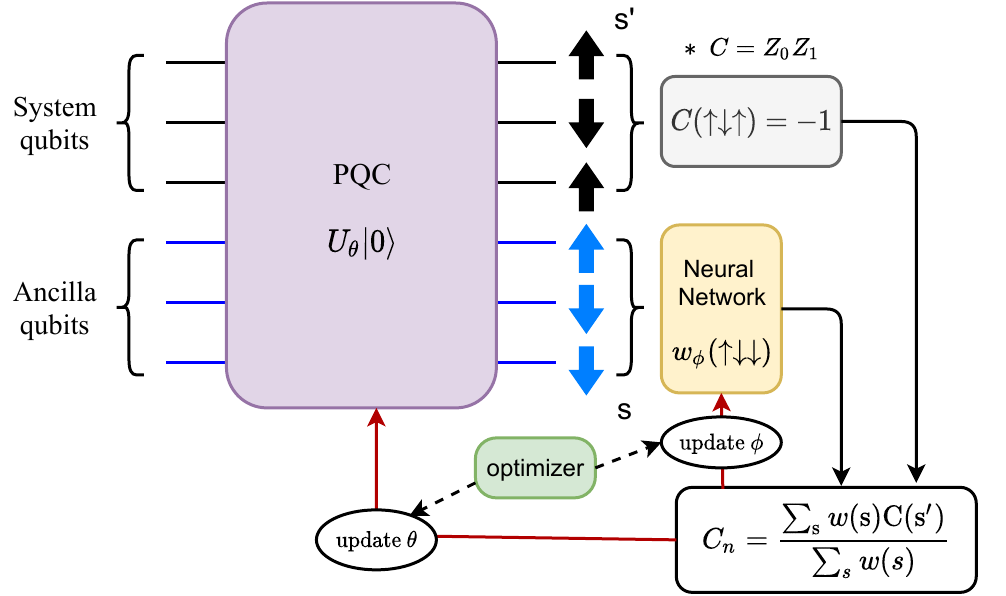}
    \caption{Schematic workflow on thermal state preparation equipped with the variational-neural post-selection scheme. The measurement outputs from ancilla qubits are reweighted by a neural network, namely, the expressive power of the ansatz is greatly enhanced with the help of the neural network and variational post-selection.}
    \label{fig:gibbs-post}
\end{figure}

    In terms of mixed state preparation such as thermal Gibbs state simulation, the ancilla qubits become more important as the measurement outputs from these qubits provide a natural implementation of the probability ensemble. In this scenario, the variational post-selection can be further combined with neural network reweighting, and the variational scheme for the mixed state preparation is shown in Fig.~\ref{fig:gibbs-post}. In this case, the post-selection is `soft', in the sense that each measurement output is kept with some neural reweighting.
    
    Overall, the post-selection idea can be regarded as a trade-off between space (the number of qubits) and time (the depth of the circuits). The complexity theory implies that one can solve a much larger set of problems if equipped with overhead-free post-selection, i.e. PostBQP=PP \cite{Aaronson2004}. Though post-selection always comes with failure overhead in practical quantum computational models, we aim to achieve better expressiveness by introducing post-selection with acceptable overhead.   In the current generation of quantum hardware, the short coherence time severely limits the depth of parameterized quantum circuits for useful quantum computation. The short-depth dilemma is even more pronounced than the limited availability of the number of usable qubits in a circuit. Therefore, it is worth trading off the circuit depth with more ancilla qubits and trial times while maintaining expressiveness. Besides, the introduction of post-selection naturally renders an effectively non-unitary quantum operation on system qubits, and the expressive capacity for a non-unitary operation is in general expected to be better than conventional unitary ansatz.

    \subsection{Results on ground state preparation}

    From a theoretical perspective, for the VQE with ancilla qubits, the output wavefunction is given by $\ket{\psi} = \sum_{ij} c_{ij} \ket{i}_s\ket{j}_a$ where $i, j$ run over system qubits $i$ and ancilla qubits $j$, respectively and $c$ is the corresponding wavefunction amplitude. For the VQE with post-selection onto the $k$-th state, the wavefunction on system qubits is reduced to $\vert \psi_k\rangle = \frac{1}{\sqrt{\sum_i \vert c_{ik}\vert^2}}\sum_i c_{ik}\vert i\rangle_s$. It is easy to identify that the energy estimation before post-selection $\langle H\rangle_\psi$ and after post-selection $\langle H\rangle_{\psi_k}$ has the following relation:
    \eq{\langle H\rangle_{\psi} = \sum_j\omega_j \langle H\rangle_{\psi_j},}{}
    where $\omega_j = \sum_i \vert c_{ij}\vert ^2$ can be taken as a probability weight since $\sum_j \omega_j = 1$ due to the normalization condition of the wavefunction. Since the average energy estimation from different post-selection is the same as the one from the VQE of no post-selection, there exists some post-selection basis $k$ such that $\langle H\rangle _{\psi_k}\leq \langle H\rangle_{\psi}$, and such $k$ basis is expected to be automatically identified by the introduction of variational module $V(\phi)$ on the ancilla qubits. This observation guarantees that we have consistently better results for the VQE estimation with the post-selection plugins.

    We report numerical results obtained via TensorCircuit \cite{Zhang2023} on different models.
    We first consider a 2D transverse field Ising model with periodic boundary condition (PBC) as the VQE target Hamiltonian to demonstrate the improvement of post-selection VQE. The Hamiltonian reads $H = \sum_{\langle i j \rangle}Z_iZ_j -\sum_i X_i$, where $\langle i j \rangle$ is the nearest neighbor sites on the square lattice with $4\times 3$ qubits. The circuit ansatz layout without ancilla qubits and post-selection scheme is given as:
    \eq{U=\prod_{p=1}^{P}\left(\prod_{i=0}^{N-1}e^{i \theta_{p i 2} X_{i}} \prod_{i=0}^{N-1} e^{i \theta_{p i 1} Z_{i} Z_{i+1}}\right) \prod_{i=0}^{N-1} H_{i},}{eq:qaoa}
    where we first apply one layer of Hadamard gates $H_i$ and then apply $P$ blocks of the $R_{ZZ}$ layer and $Rx$ layer repetitively.
    The two-qubit gate layout for the ZZ layer is arranged in a ladder fashion as (0, 1), (1, 2) $\cdots$.
    We only introduce one ancilla qubit in the variational post-selection case, labeled as the n-th qubit, and the variational post-selection module is a parameterized $SU(2)$ rotation with the post-selection on the computational basis $\vert \uparrow\rangle$. Accordingly, the two-qubit gate layout changes with ancilla qubits and is composed of an `all-to-one' pattern as (0, n), (1, n) $\cdots$. It is worth noting that the new layout consumes merely the same amount of quantum resources compared to the ladder layout in quantum processors that contain a qubit subset of 1D ring topology (see the Supplemental Materials for detailed quantum resource analysis and comparison).
    The VQE energy estimations are compared in Table.~\ref{tab:tfimvqe} where the post-selection scheme holds noticeable advantage. Besides, the overall success probability for post-selection is in general in the range of $[0.5, 0.7]$ which avoids exponential small successful probability due to the small number of post-selection qubits. 

\begin{table}[]
	\caption{VQE results on 2D TFIM energy estimation with and without post-selection. The exact ground state energy is $-18.914$ given by exact diagonalization.}
	\begin{tabular}{@{}cccc@{}}
		\toprule
		circuit depth (P)  & 2      & 3      & 4      \\ \midrule
		conventional VQE    & -14.81 & -15.41 & -15.62 \\ \midrule
		VQE with post-selection & -18.59 & -18.67 & -18.80 \\ \bottomrule
	\end{tabular}
\label{tab:tfimvqe}
\end{table}

\begin{figure}[t]\centering
    \includegraphics[width=0.5\textwidth]{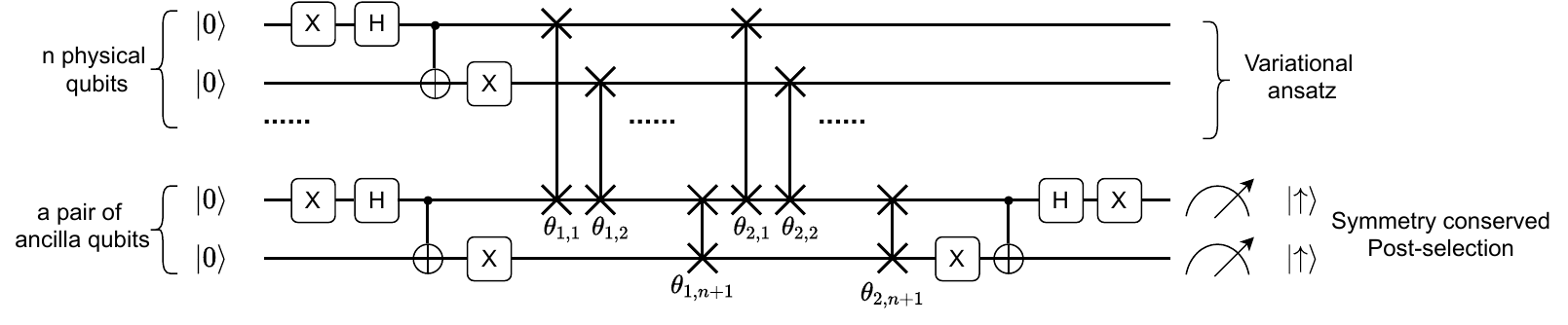}
    \caption{$SU(2)$ symmetry enforced VQE ansatz with post-selection enabled. Besides the symmetric initial state and the symmetric circuit ansatz, the post-selection must also keep $SU(2)$ symmetry.}
    \label{fig:sym-ansatz}
\end{figure}

    We also employ a 2D PBC Heisenberg model on the square lattice, whose Hamiltonian is given by $H=\sum_{\langle ij\rangle} X_iX_j + Y_iY_j + Z_iZ_j$, where $\langle ij\rangle$ represents nearest neighboring sites on the lattice. This model is more interesting as it hosts intrinsic $SU(2)$ symmetry and such symmetry can be exploited by the VQE to boost the ground state simulation. We design a symmetry-enforced post-selection scheme compatible with symmetric VQE ansatz. In conventional VQE, the $SU(2)$ symmetry is kept by the Bell pair initial state and the circuit ansatz composed of parameterized SWAP $R_{swap}$ gate $e^{i\theta \text{SWAP}}$. To keep this symmetry with ancilla qubits and post-selection, we introduce an even number of ancilla qubits so that we still have nontrivial $J_{tot}^2=0$ sector, and the overall ansatz is depicted in Fig.~\ref{fig:sym-ansatz}. Basically, we make one ancilla qubit to interact with each system qubit sequentially and symmetrically (with respect to $SU(2)$ symmetry). After each round of the ansatz, the two ancilla qubits also go through an $R_{swap}$ gate. In the symmetry-enforced case, the post-selection is now fixed and effectively projecting out a Bell pair since only such post-selection can keep the state of physical qubits in the $J_{tot}^2=0$ sector. With such ansatz, the results of the Heisenberg model VQE on $4 \times 3$ square lattice are summarized in Table.~\ref{tab:heisenberg-vqe}, where conventional VQE is made of $R_{swap}$  in ladder layout and symmetry-breaking post-selection scheme post select computational basis $\vert \downarrow \downarrow\rangle$ on the pair of ancilla qubits. The results indicate the importance of symmetry enforcement in the VQE as well as in the post-selection module.

\begin{table}[]
    \caption{VQE results for 2D Heisenberg model on $4\times 3$ qubits. The exact ground state energy is $-29.473$. We find that the symmetry-enforced post-selection scheme gives the best estimation for ground state energy.}
    \begin{tabular}{@{}cccc@{}}
    \toprule
    circuit depth (P)  & 2      & 3      & 4      \\ \midrule
    conventional VQE    & -25.57 & -28.29 & -28.85 \\ \midrule
    VQE with ancilla qubits&&&  \\ (symmetric post-selection) & -25.8 & -28.36 & -29.05 \\ \midrule
    VQE with ancilla qubits&&&\\ (symmetry-breaking post-selection)& -20.98&-23.49& -24.78\\ \midrule
    VQE with ancilla qubits&&&  \\ (no post-selection)&-24.27 & -26.65 &  -28.50  \\ \bottomrule
    \end{tabular}
    \label{tab:heisenberg-vqe}
\end{table}

\begin{figure}[t]\centering
    \includegraphics[width=0.4\textwidth]{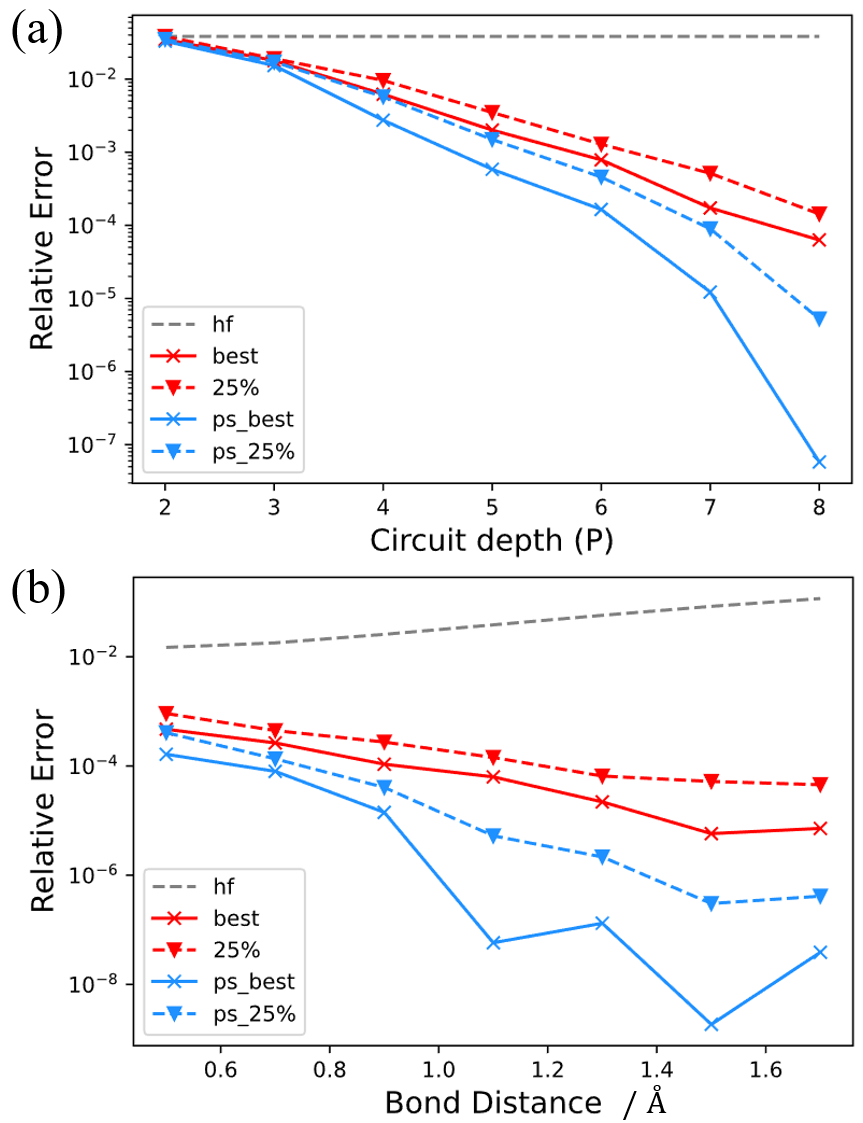}
    \caption{Results for H$_4$ chain molecule. (a) Relative energy error for VQE with respect to the circuit depth ($P$) when the bond distance is $1.1 \mathrm{\AA}$ between neighboring H atoms. (b) Relative energy error for VQE with respect to the bond distance when the circuit depth $P = 8$.
    The red lines represent the results of the standard VQE and the blue ones represent the results of the VQE with post-selection on one ancilla qubit. The lines with crosses represent the best result among the 50 optimization trials of different initialized parameters, while the dashed lines with inverted triangles represent the best 25-percentile optimization results. The grey dashes indicate the Hartree-Fock energies.}
    \label{fig:h4_chain}
\end{figure}

    The post-selection VQE method can be also applied to molecular systems. Due to the conservation of electron number for a given molecule model, we employ a hardware-efficient ansatz incorporating $U(1)$ symmetry with Jordan-Wigner transformation. For the post-selection scheme, we predefine the post-selection output in $\vert\uparrow\rangle$ basis, such that the $U(1)$ symmetry (or the conservation of electron number) is retained.
    The ansatz (including the variational quantum module $V$) is constrained to $R_z$, $R_{zz}$, and $R_{swap}$  gates to maintain the $U(1)$ symmetry. Besides, some X gates are added at the beginning of the ansatz to prepare a Hatree-Fock initial state (see the SM for details).
    
    Utilizing the Hydrogen-4 chain model (see Supplemental Materials for more detailed information on the molecule models), we approximate its ground-state energies using the VQE with and without post-selection. With 50 independent optimization trials from different random initialization parameters, we numerically obtain results with circuit depth from $P = 2$ to $P = 8$ at a bond distance of $1.1 \mathrm{\AA}$ (see Fig. \ref{fig:h4_chain}(a)) and with each bond distance from $0.5$ to $1.7 \mathrm{\AA}$ at the circuit depth of $P = 8$ (see Fig. \ref{fig:h4_chain}(b)). From the results, we identify orders-of-magnitude reduction in relative error after using the post-selection scheme. 

\begin{figure}[t]\centering
    \includegraphics[width=0.45\textwidth]{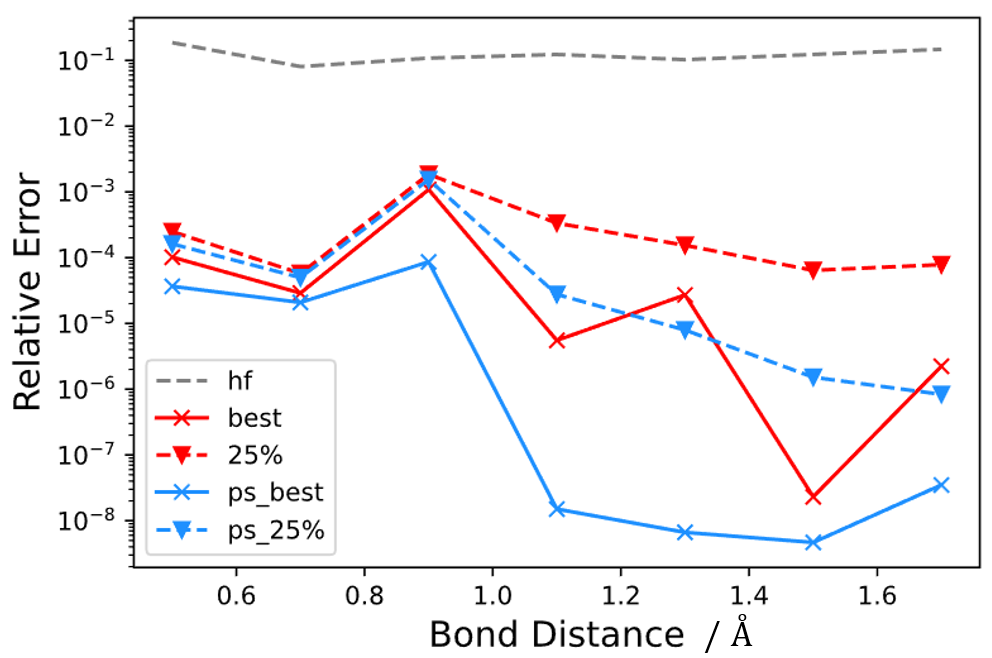}
    \caption{Results for H$_4$ square model. Relative energy error for VQE with respect to the bond distance when the circuit depth $P = 8$.
    The red lines represent the results of the standard VQE and the blue ones represent the results of the VQE with post-selection. The lines with crosses represent the best result among the 50 optimization trials of different initialized parameters, while the dashes with inverted triangles represent the best 25-percentile results. The grey dash indicates the Hartree-Fock energies as a baseline.}
    \label{fig:h4_square}
\end{figure}

\begin{figure}[t]\centering
    \includegraphics[width=0.4\textwidth]{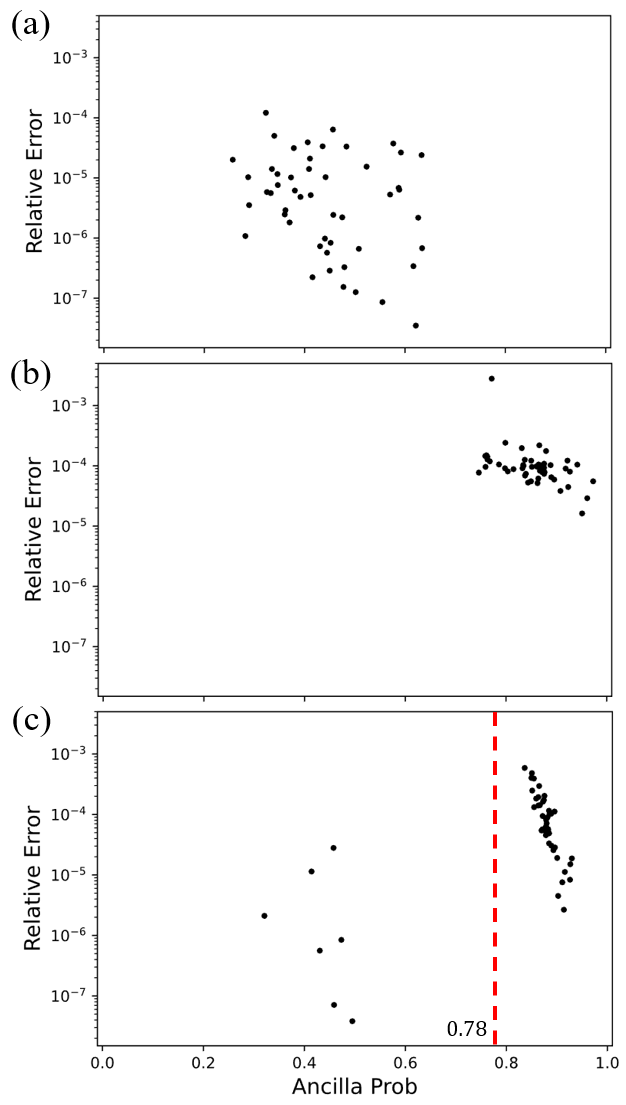}
    \caption{Ancilla post-selection success probabilities, with the test bed H$_4$ square model. (a)-(c) Relative error with respect to the success probabilities of post-selection. (a) The objective is the expectation of the Hamiltonian without ancilla success probability. (b) The objective is Eq. \ref{eq:loss1}. (c) The objective is Eq. \ref{eq:loss2} with a sigmoid ancilla processing ($p_0 = 0.78$).}
    \label{fig:h4_square_diffCost}
\end{figure}

    We further applied the post-selection scheme to the H$_4$ square model to test its effectiveness in molecules having higher spatial dimensions. We report the numerical results by fixing the circuit depth to $P = 8$ and varying the bond distance from $0.5$ to $1.7 \mathrm{\AA}$. From the results shown in Fig. \ref{fig:h4_square}, an orders-of-magnitude decrease in the relative error is also witnessed, showcasing the universal capability of our proposed method.

    Additionally, we explore redesigning the objective function to minimize the post-selection overhead. We use the H$_4$ square model at the bond distance of $1.7 \mathrm{\AA}$ with VQE circuit depth $P=8$ as the testbed. Previously, we employed the expectation value of the model Hamiltonian under post-selected wavefunction $\vert \psi_k\rangle$, $\langle H\rangle _{\psi_k}$, as the objective function for optimizing the VQE. The success probabilities of post-selecting the ancilla qubit are shown in Fig. \ref{fig:h4_square_diffCost}(a). The results indicate that the correlation between post-selection success probability and the energy estimation accuracy is relatively small. That is to say, we might reduce the post-selection overhead (failure probability) while maintaining energy accuracy by engineering suitable objective functions. Therefore, we introduce a new objective function:
    \eq{Obj_1 = \langle H\rangle _{\psi_k} - \lambda \times p_a , }{eq:loss1}
    where $\lambda$ is a positive penalty hyperparameter and $p_a$ is the success probability of post-selecting the ancilla qubit. However, the relative energy error increases when optimizing such an objective function (see Fig. \ref{fig:h4_square_diffCost}(b)). Consequently, we design the objective function by choosing the sigmoid function to filter the success probability of the post-selection as 
    \eq{Obj_2 = \langle H\rangle _{\psi_k} - \lambda ~ \mathrm{sigmoid}(p_a - p_0) , }{eq:loss2}
    where $p_0$ is the hyperparameter for the offset of the sigmoid ancilla processing (see the SM for details). The optimized result under such an objective function is shown in Fig. \ref{fig:h4_square_diffCost}(c).

    \subsection{Results on thermal state preparation}
    We now report the results on thermal state preparation with variational-neural post-selection as shown in Fig.~\ref{fig:gibbs-post}. We further compare our results with previous common variational pipelines for Gibbs state approximation.
    
    The difficulties in preparing high-fidelity thermal states are not only in the design of variational ansatz but are also related to the experimental protocol for evaluating objective functions. In principle, we need to minimize the free energy to approximate the thermal equilibrium state, which is given by $F_G(\rho)=\rm{Tr}(H\rho)-\frac{1}{\beta}(-\rm{Tr}(\rho\ln \rho))$, where $\beta=1/T$ is the inverse temperature. However, the second term for von Neumann entropy is nonlinear with $\rho$, and in general requires exponential times of measurements to evaluate accurately. Previous approaches \cite{Wang2020b, Chowdhury2020} utilize the truncated series expansion of the entropy term so that only finite terms like $\rm{Tr}(\rho^n)$ need to be evaluated which can be obtained efficiently (polynomial scaling regarding the system size) via randomized measurements \cite{VanEnk2011, Brydges2019} or SWAP test \cite{Abanin2012, Daley2012, Foulds2020}. However, the accuracy for such truncated series approximation, though guaranteed in theory, is very resource-demanding to maintain as it requires lots of quantum resources for evaluating $\rm{Tr}(\rho^n)$ when $n$ is large. The other route is to utilize the classical neural probability model to learn the probability distribution of thermal equilibrium systems whose entanglement can be easily obtained by classical calculation \cite{Verdon2019, Liu2019}. Though this approach renders the free energy objective tractable, the expressiveness is much more restricted compared to variational ansatz with ancilla qubits when the number of quantum gates is the same (see Supplemental Materials for details on settings from previous works and numerical comparisons).
    
    In this work, we combine the variational circuit ansatz with neural reweighted post-selection on ancilla qubits and utilize a novel objective function, Renyi free energy: 
    \eq{F_\alpha (\rho) = \rm{Tr}(H\rho) -\frac{1}{\beta_\alpha}\left(\frac{1}{1-\alpha}\ln \rm{Tr}(\rho^\alpha)\right)}{arg2}
    The minimization of this objective leads to the so-called Renyi thermal state instead of the Gibbs thermal state. For simplicity, we use the second order $\alpha=2$ Renyi entropy throughout the work. The corresponding $\alpha=2$ Renyi state, different from Gibbs state of distribution $p_i\propto e^{-\beta E_i}$ in eigen-space, follows a linear distribution as $p_i\propto \rm{max}(0, 1-\frac{\beta}{2}(E_i-\bar{E}))$, where $\bar{E} = \rm{argmin}_{\bar{E}} F_2(\rho(\bar{E}))$. Although the ensemble distributions for Gibbs state and Renyi state look very different, as demonstrated in \cite{Giudice2021, Lu2024}, Renyi thermal state gives consistent local observable expectations as corresponding Gibbs thermal states in the thermodynamic limit. Therefore, by utilizing Renyi free energy as the objective function, we make a good trade-off between the efficient evaluation of the objective and the accuracy of local property predictions for the thermal equilibrium system.

    Via the variational-neural post-selection setup in Fig.~\ref{fig:gibbs-post}, the output state from the circuit is $\ket{\psi}=U\ket{0_s}\ket{0_a}$, by defining the amplitude as $\psi_{im} = \bra{i_s}\bra{m_a}\ket{\psi}$ and after the neural post-selection, we have the target mixed state on system qubits as
    \eq{\rho_{ij} = \sum_m \psi_{im} f(m)\psi^*_{jm},}{}
    where the sum is over computational basis $m$ on ancilla qubits and $f$ is the neural network. After evaluating both the energy and Renyi entropy terms on the quantum computer, parameters from the circuit and neural network are optimized accordingly to minimize Renyi free energy. The evaluation of the observable is done by so-called neural reweighting, namely, we collect several measurement shot results as bitstring $sa$, and the observable value for each shot is defined as $C(s)$. Then the final averaged correlation evaluation is 
    \eq{\overline{C} = \frac{\langle f(a)C(s)\rangle_{sa}}{\langle f(a)\rangle_{sa}},}{}
    and the denominator is here to normalize the neural weights $f(a)$ as a probability distribution. 
    The previous variational Gibbs state preparation scheme corresponds to the case where $f$ is a constant: $f(a)\equiv 1, \forall a$.
    If the neural network output range is too large, the estimation fluctuation on the correlation is also large which leads to a higher number of required measurements (similar to the case investigated in \cite{Zhang2021b}). Therefore, we could restrict the output range of the neural network $f$ to reduce the required number of measurements for a given estimation precision and we call this extension bounded neural reweighting method. As we show from the numerical results below, bounded neural reweighting has a similar or even better performance compared to unbounded neural reweighting methods.

    We benchmark our new variational thermal state preparation scheme on a 1D TFIM system (N=8). We introduce the same number of ancilla qubits as system qubits and make them interleave with each other. We use the variational ansatz as given by \Eq{eq:qaoa} on the 16-qubit system. We utilize fidelity with exact Gibbs state and local Pauli observable expectations as indicators for the quality of thermal state preparation.  The fidelity between two mixed states $\rho$ and $\rho_0$ is defined as \eq{F = \left(\operatorname{Tr}\left(\sqrt{\sqrt{\rho_{0}} \rho \sqrt{\rho_{0}}}\right)\right)^{2}.}{}
    
\begin{figure}[t]\centering
    \includegraphics[width=0.5\textwidth]{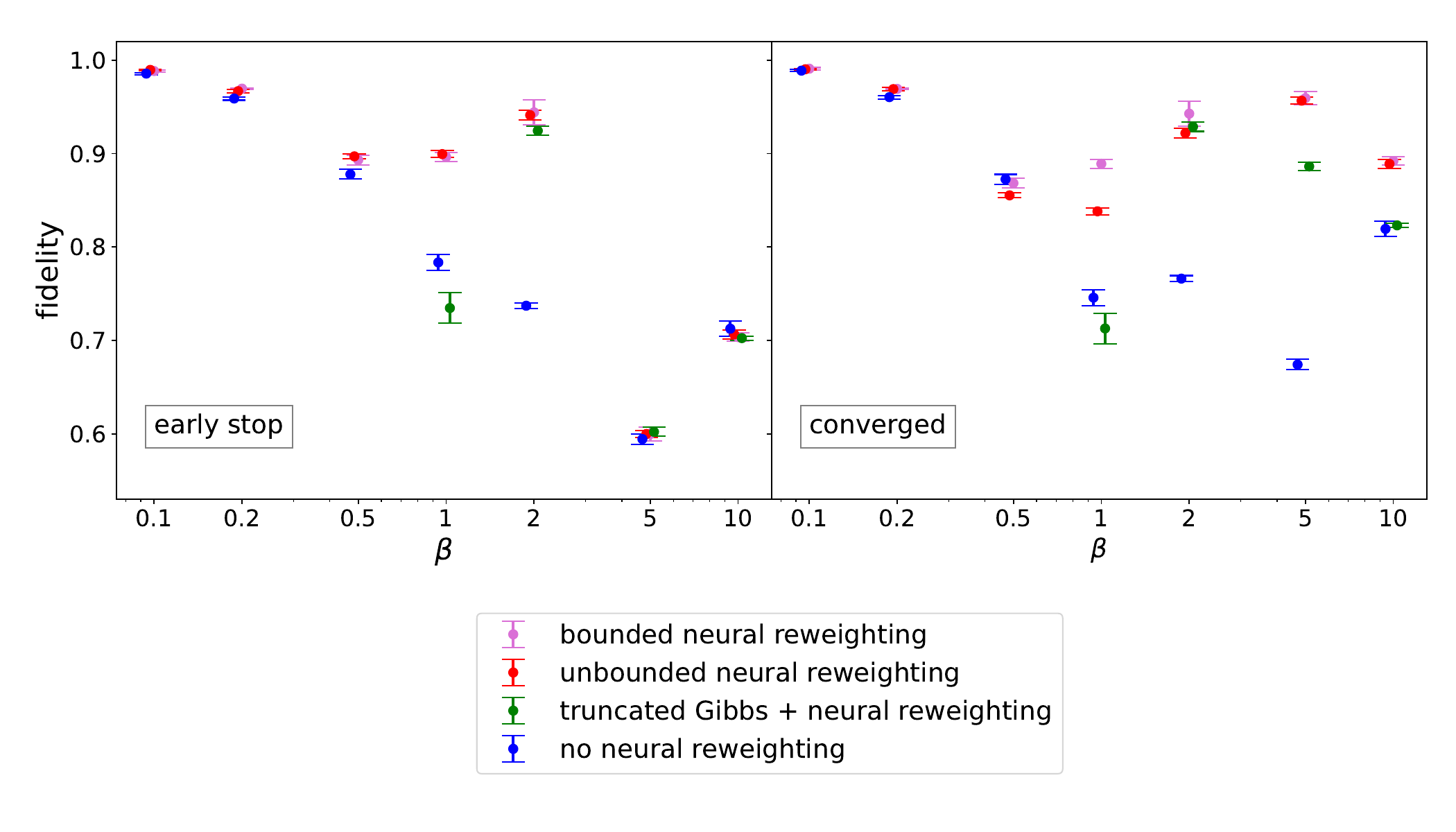}
    \caption{The fidelity between variational state and exact Gibbs state with different inverse temperature $\beta$. Early stop results are recorded after 300 rounds of gradient descent optimization while converged results are after 2600 rounds of gradient descent optimization.}
    \label{fig:fidelity}
\end{figure}

\begin{figure}[t]\centering
    \includegraphics[width=0.5\textwidth]{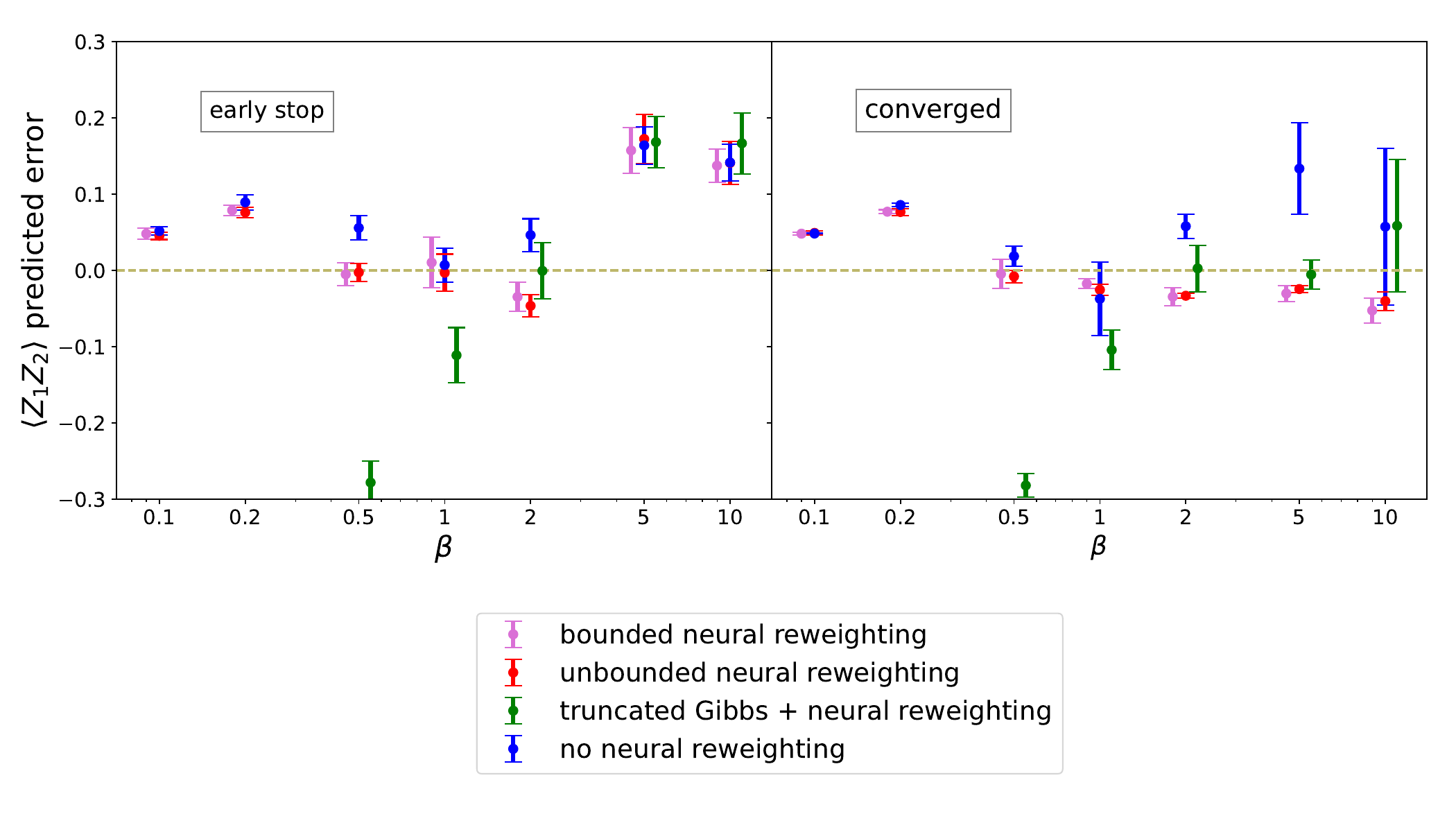}
    \caption{Correlation prediction error away from exact correlation values (We use $\langle Z_1Z_2\rangle $ here, other local correlations give qualitatively similar results). Early stop results are recorded after 300 rounds of gradient descent optimization while converged results are after 2600 rounds of gradient descent optimization.}
    \label{fig:corr}
\end{figure}

    From the results in Fig.~\ref{fig:fidelity} and Fig.~\ref{fig:corr}, we find that early stop in optimization often has better results for high-temperature regions while converged results are more consistent for low-temperature regions.
    Besides, while the bounded neural reweighting approach seems to have worse expressive power, it gives similar or even better results than unbounded neural reweighting while requiring a much smaller number of measurements. The gains of bounded neural reweighting and early stop in training are because the target we utilize is Renyi free energy, and overfitting to this objective may drive the results away from the thermal equilibrium Gibbs state. The advantage of the choice of the objective function is made clear by comparison with the converged results by minimizing second-order Taylor expansion of Gibbs free energy, which we call truncated Gibbs in the figure. The results obtained by optimizing Taylor truncated Gibbs loss are much worse than the one from Renyi free energy.
    
    Moreover, we emphasize that the deviation from the correct correlation will be further suppressed for the large-size system since the equivalence between the Renyi state and the Gibbs state in terms of local properties is guaranteed in the thermodynamic limit \cite{Giudice2021}.

    In summary, equipped with our newly proposed neural reweighting ansatz as well as the Renyi free energy as the objective function, we achieve good results for thermal equilibrium state preparation on quantum computers where high-quality estimation of local properties can be obtained.

\section{Discussion}
    In this work, we introduce the post-selection scheme but have only evaluated it with a modest number of ancilla qubits and a relatively simple circuit ansatz. However, the results herein are still encouraging. It is an interesting future direction to explore more suitable circuit architectures with more ancilla qubits for given tasks where automatic quantum architecture search algorithms \cite{Zhang2020b, Du2020a, Lu2020, Zhang2021neural} may play a crucial role. The inspiration from imaginary time evolution or space-time duality \cite{Ippoliti2020a} from quantum physics may also be relevant. 

    It is also straightforward to incorporate the scalable neural post-processing on system qubits \cite{Zhang2021b, Zhang2021d} together with our variational post-selection schemes on ancilla qubits which can further enhance the capability of VQAs by adopting more power from neural networks in different modules.

    The introduction of post-selection gadgets can also be instrumental to quantum noise resilience. The post-selection scheme is also directly compatible with several quantum error mitigation \cite{Cai2023} schemes such as symmetry verification \cite{Bonet-Monroig2018, Sagastizabal2019}. It is an interesting future direction to study the interplay between quantum noises and variational quantum algorithms with post-selection.

~\newline

%

\clearpage
\newpage

\begin{widetext}
    \section*{Supplemental Materials}
    \renewcommand{\theequation}{S\arabic{equation}}
    \setcounter{equation}{0}
    \renewcommand{\thefigure}{S\arabic{figure}}
    \setcounter{figure}{0}
    \renewcommand{\thetable}{S\arabic{table}}
    \setcounter{table}{0}
    
    \subsection{Quantum resource analysis for post-selection scheme with quantum compiling on real hardware topology}
    
    The variational post-selection ansatz we use is to make one ancilla qubits interact with all other system qubits sequentially.  At first sight, one layer of such ansatz is much more expensive than the vanilla ansatz with two-qubit gates in a ladder layout by considering a 1D topology for the quantum hardware. However, we demonstrate that the quantum resources required for one layer of the post-selection ansatz are nearly the same as the vanilla ansatz in the 1D ring topology, which justifies the performance comparison in the main text. It is worth noting that the quantum hardware in a 1D ring topology is common as a subpart from a large 2D qubit array, namely, our analysis here directly applies to real superconducting quantum processors. We believe that with further efforts, more ansatz of better expressiveness and hardware compatibility can be identified.
    
    Considering the two-qubit gate layout in the periodic boundary condition, the number of general two-qubit gates is the same as the system size $N$. Now consider the ansatz with the introduction of one extra ancilla qubit, and make this qubit couple with each system qubit by two-qubit gates sequentially. In this case, the basic idea is to exchange the ancilla qubit and the system qubit after the application of each two-qubit gate. Since the swap gate and the two-qubit parameterized gate are applied on the same pair of qubits at the same time, they can be merged as one general two-qubit gate and recompile to the hardware gate set. In this way, the ancilla qubit goes around the ring for one layer ansatz, see Fig. \ref{fig:ring} for an illustration. In this way, the total number of general two-qubit gates required in the post-selection scheme is $N+1$, which is nearly the same as the original VQE ansatz. Therefore, our comparison between these two types of variational ansatz is fair, and post-selection indeed enhances the expressiveness with the same amount of quantum resources.
    
    For the symmetry-enforced post-selection scheme, the same logic applies here. Though the ansatz we utilized in the Heisenberg model VQE case requires a pair of ancilla qubits, only one of the ancilla interacts with all other qubits (system qubits and the other ancilla qubits). Therefore, for each layer of the ansatz, this ancilla qubit goes around the ring topology with the help of the swap gate, and the required number of general two-qubit gates is still in the order of the system size, comparable with the conventional VQE with the same number of layers.

\begin{figure}[b!]\centering
    \includegraphics[width=0.5\textwidth]{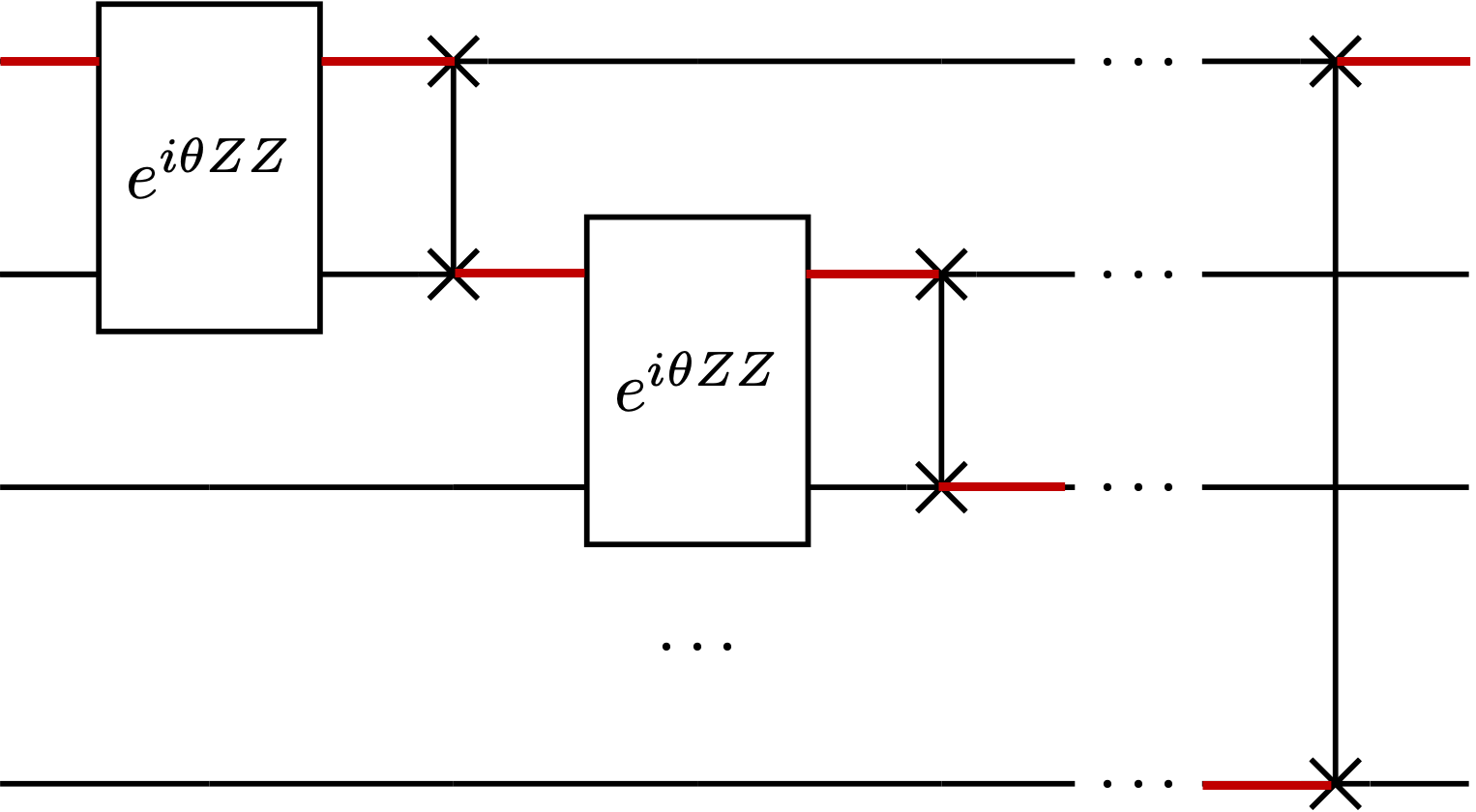}
    \caption{The schematic circuit ansatz implementation on quantum hardware of one-dimensional ring topology. In real experiments, swap gates can be merged with the parameterized two-qubit gates before and recompiling as a general two-qubit gate. The red line indicates the position of the ancilla qubit, which goes around the hardware for one layer of variational ansatz with the post-selection scheme.}
    \label{fig:ring}
\end{figure}

    \subsection{The molecule models and the particle number conserving circuit ansatz}

    In the main text, we use the H$_4$ chain and the H$_4$ square for simulation. The H$_4$ chain is four equispaced Hydrogen atoms in a line. When the bond distance changes, the distances between the adjacent Hydrogen atoms change accordingly. The H$_4$ square is four Hydrogen atoms forming a square. The basis set used by both the molecules is the STO-3G basis set and the Jordan-Wigner transformation is applied for fermion qubit mapping. After the transformation, the two H$_4$ models have 8 qubits (9 for post-selection).

    In terms of the hardware-efficient ansatz with $U(1)$ symmetry, we use the ansatz shown in Fig. \ref{fig:keep_u1}, which satisfies the requirements of only $R_z$, $R_{zz}$, and $R_{swap}$ gates are used in the ansatz. To introduce the initial electron occupation number, X gates, equal in number to the electrons in the molecule model, are applied at the beginning of the whole ansatz to obtain the Hartree-Fock initial state.

    We explore redesigning the objective function to minimize the post-selection overhead in the main text. We add the sigmoid function as the ancilla probability processing term to the objective function  
    \eq{Obj_2 = \langle H\rangle _{\psi_k} - \lambda ~ \mathrm{sigmoid}(p_a - p_0) , }{}
    where $p_0$ is the offset of the sigmoid ancilla processing. Here, we give an example with $p_0 = 0.78$ shown in \ref{fig:supp_sigmoid}. We choose the sigmoid function as the ancilla processing since its non-linear nature gives the post-selection with a low success probability greater punishment such that the post-selection overhead could be greatly reduced. The sigmoid function used here is a rather simple mapping but still works well in minimizing the post-selection overhead while maintaining accuracy. Therefore, we believe that with a more complicated and proper ancilla processing method, the post-selection overhead problem can be better resolved.

\begin{figure}[t!]\centering
    \includegraphics[width=0.95\textwidth]{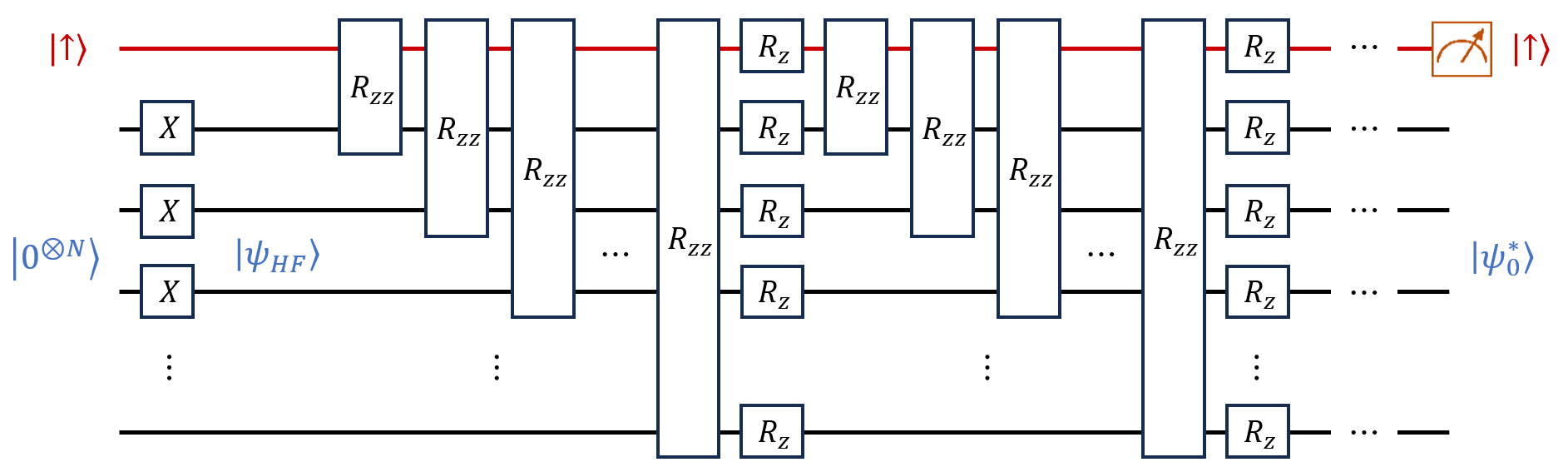}
    \caption{The hardware-efficient ansatz with $U(1)$ symmetry. X gates are first applied to the corresponding electron orbitals of the HF state in the front of the ansatz, thus turning the initial state $\ket{\bf 0}$ into the Hartree-Fock state $\ket{\psi_{HF}}$. Then follows the $U(1)$ symmetry circuit (2 blocks are shown here), with only $R_z$, $R_{zz}$, and $R_{swap}$ gates.}
    \label{fig:keep_u1}
\end{figure}

\begin{figure}[t]\centering
    \includegraphics[width=0.35\textwidth]{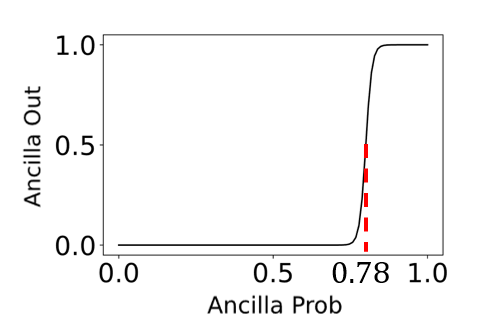}
    \caption{The sigmoid ancilla processing.}
    \label{fig:supp_sigmoid}
\end{figure}

    \subsection{Expressiveness comparison between neural pre-processing scheme and neural post-processing scheme} 
    
    {\bf The neural pre-processing approach.}
    As we have mentioned in the main text, the entropy term in the free energy in general requires an exponential number of measurements to reliably estimate, and this is why we instead use Renyi free energy as an alternative. There is another scheme which we call the neural pre-processing approach, avoiding the entropy estimation problem. The basic setup is as follows. Firstly, we have a neural model for the probability distribution of classical bitstrings, i.e. by providing $s\in \{0, 1\}^n$, we have a parameterized classical function that gives $P_\phi(s)$ as a probability distribution. The probabilistic models commonly used are restricted Boltzmann machine or variational autoregressive models. We first sample bitstring $s$ from such probabilistic model $P$ and feed the corresponding direct product state $\vert s\rangle$ to a parameterized quantum circuit $U_\theta$, which gives the final state $\vert \psi_s\rangle= U_\theta \vert s\rangle$. Therefore, the output mixed state is defined as 
    \eq{\rho= \sum_s P(s)\vert \psi_s\rangle\langle \psi_s\vert.}{}
    Since the states $\vert \psi_s\rangle$ are orthogonal with each other, the quantum Gibbs entropy can be reduced to the classical one as 
    \eq{S = -\rm{Tr}\left(\rho \ln \rho\right) = -\rm{Tr}\left(P(s)\ln P(s)\right).}{}
    Therefore, we can efficiently evaluate Gibbs entropy in this setup,  avoiding the exponential number of measurement shots. This neural pre-processing approach seems promising as it also combines the capability of classical neural modules, provides efficient estimation of exact Gibbs free energy, and even requires no ancilla qubits. However, by numerical experiments below, we demonstrate that such a neural pre-processing approach has more limited expressiveness compared to our neural post-processing approach. Intuitively, the PQC $U$ in the neural pre-processing approach is equivalent to the diagonalization transformation matrix, which requires a very deep circuit to implement for high-quality approximations.
    
    We now describe our numerical experiment setup for the comparison. Again, we use 8-site TFIM thermal state preparation as the testbed. For this problem, our approach requires ancilla qubits, and we use the same number of ancilla qubits as the system qubits and hence run the experiments on 16-qubit circuits. To make this a fair comparison, we use a more powerful variational ansatz for neural pre-processing approach as 
    
    \eq{U=\prod_{p=1}^{P}\left(\prod_{i=0}^{N-1}e^{i \theta_{p i 2} Y_{i}}e^{i \theta_{p i 2} X_{i}} \prod_{i=0}^{N-1}e^{i \theta_{p i 1} Z_{i} Z_{i+1}}\right) \prod_{i=0}^{N-1} H_{i},}{eq:qaoaplus}
    where we further introduce one layer of Ry rotation for each repetition block. For our neural post-selection scheme, we still stick to the simpler ansatz as given in the main text. Besides, since our scheme costs twice the number of qubits, we make a comparison between the neural pre-processing scheme with depth $2P$ and our results with depth $P$. Since we care about the expressiveness of these two approaches, we directly use Gibbs free energy as the objective to run the optimization numerically. The results are shown in Table.~\ref{tab:com}. Apart from fidelity, the trace distance between two mixed states $\rho$ and $\sigma$ is defined as:
    \eq{T(\rho, \sigma) = \frac{1}{2}\vert\vert \rho-\sigma \vert\vert_1 = \frac{1}{2}\rm{Tr}\left ( \sqrt{(\rho-\sigma)^\dagger(\rho-\sigma)}\right ).}{}
    
      As we can see, our scheme approximates the Gibbs state surprisingly well (fidelity reaches 0.996) with only two blocks of the ansatz (P=2). On the contrary, even with more complicated ansatz and more blocks (2P=16), the previous neural pre-processing setup still has much worse performance. 
    
    \begin{table}[t]
        \caption{Expressiveness comparison between neural pre-processing approach and neural post-processing approach (ours).  Our approach greatly outperforms the previous setting.}
        \begin{tabular}{@{}ccccc@{}}
            \toprule
            Circuit Blocks: P  & ~& 2      &  4     & 8      \\ \midrule
            \multirow{3}*{neural pre-processing (2P) }  & fidelity& 0.823 & 0.926 & 0.976 \\ \cline{2-5}
            &trace distance&0.389&0.197&0.0977\\ \cline{2-5}
            &Gibbs free energy&-10.834&-11.183&-11.301\\
            \midrule
            \multirow{3}*{neural post-processing (ours) (P)  }  & fidelity& 0.996 & 0.997 & 0.997 \\ \cline{2-5}
            &trace distance&0.058&0.051&0.051\\ \cline{2-5}
            &Gibbs free energy&-11.343&-11.347&-11.347\\
            \bottomrule
        \end{tabular}
    \label{tab:com}
    \end{table}

    \subsection{More results on thermal state preparation}
    
    Since the objective function we utilized in the optimization is in fact the Renyi free energy, we can compare the converged ensemble against the exact Renyi state, which has no overfitting issue. The final fidelity against the exact Renyi state for different temperatures is shown in Fig.~\ref{fig:renyifidelity}. Since in terms of fitting Renyi state, the training has no overfitting issue, the results get improved with more training epochs or more expressive ansatz. The results are consistent with this observation.
    
    We also show several other few-body correlation predictions in Fig.~\ref{fig:x3} and Fig.~\ref{fig:z0z7}. The conclusion remains qualitatively the same as the main text. For high-temperature regions (small $\beta$), early stop results are sufficient while more training epochs are required for low-temperature regions (large $\beta$). More importantly, neural reweighting is the only approach to obtain reliable correlation predictions at different temperatures, both truncated Gibbs loss function and plain variational ansatz (no neural reweighting) fail to give consistent correlation prediction with the exact values. Moreover, the quality of the converged results is similar for both bounded and unbounded neural reweighting schemes. Thus, to reduce the required number of measurement shots, bounded neural reweighting is preferred in practical usage.
    
\begin{figure}[t]\centering
    \includegraphics[width=0.7\textwidth]{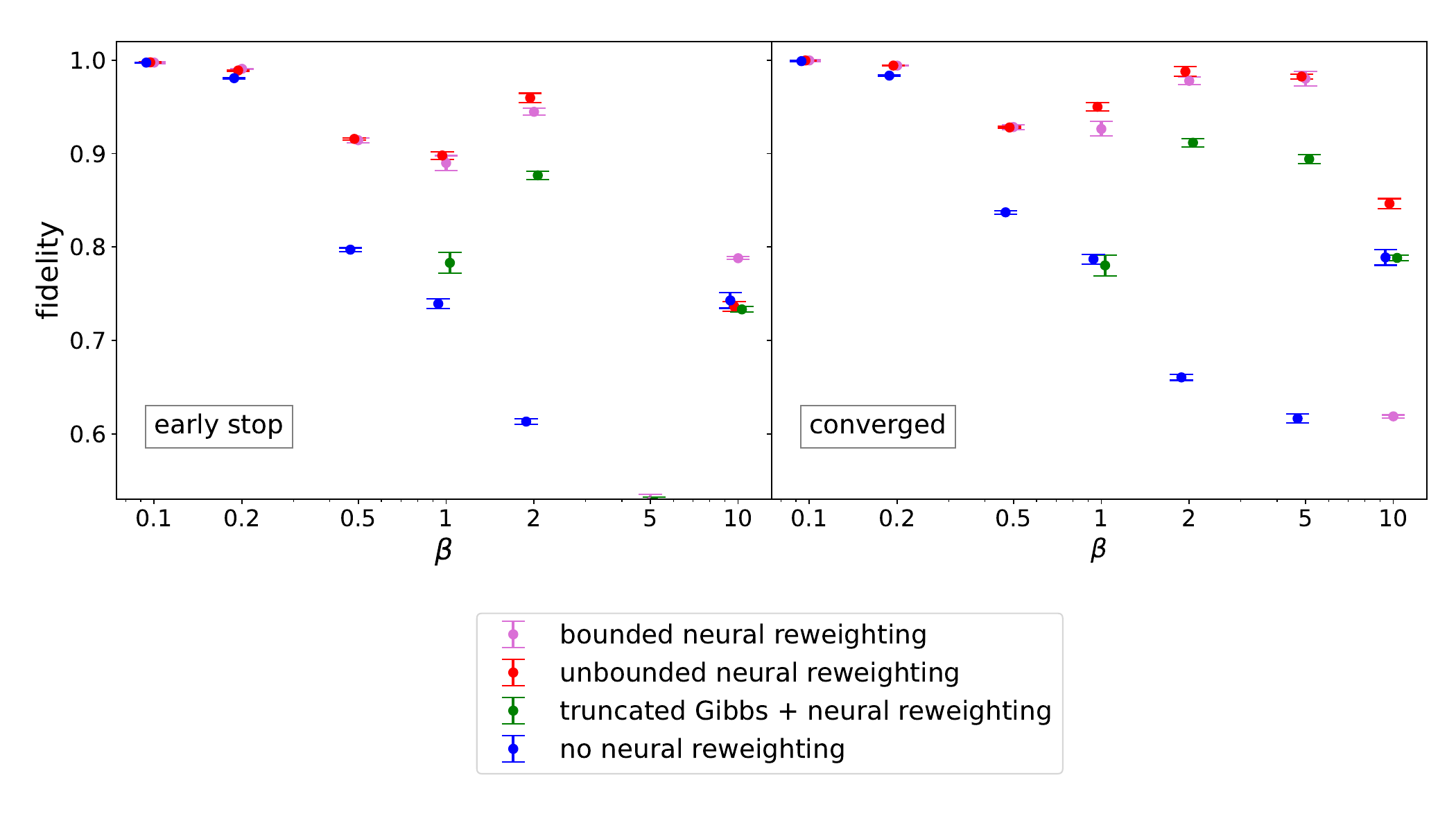}
    \caption{The fidelity between variational state and exact Renyi state. Early stop results are recorded after 300 rounds of gradient descent optimization while converged results are after 2600 rounds of gradient descent optimization. The converged results are consistently improved compared to early stop results, as there is no overfitting issue in terms of fitting the Renyi state. Besides, unbounded neural reweighting has the best results for the same reason.}
    \label{fig:renyifidelity}
\end{figure} 

\begin{figure}[t]\centering
    \includegraphics[width=0.7\textwidth]{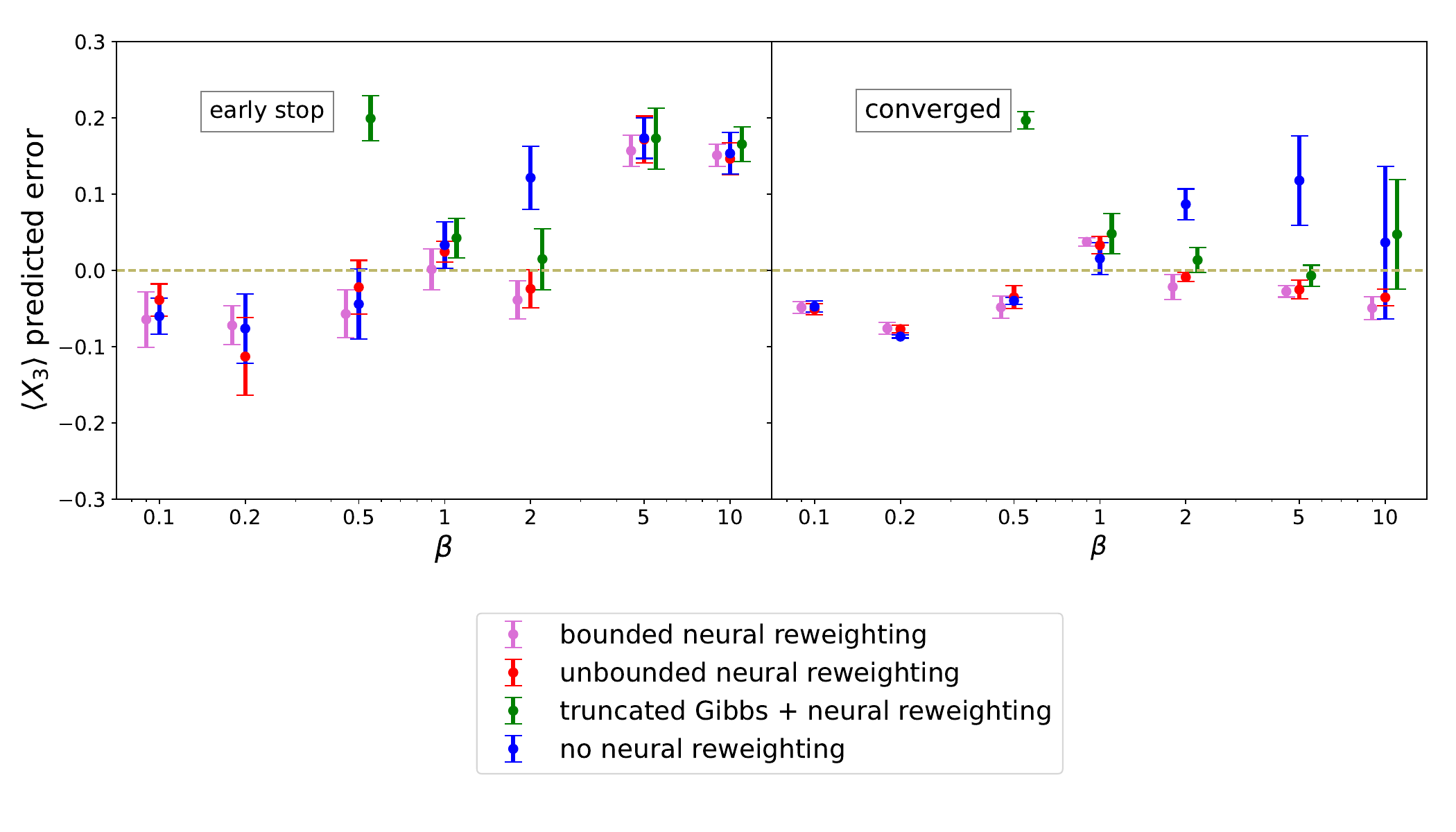}
    \caption{Correlation $\langle X_3 \rangle$ prediction error away from exact correlation values. Early stop results are recorded after 300 rounds of gradient descent optimization while converged results are after 2600 rounds of gradient descent optimization. Points beyond the scope of this figure are not shown due to large prediction errors.}
    \label{fig:x3}
\end{figure} 

\begin{figure}[t]\centering
    \includegraphics[width=0.7\textwidth]{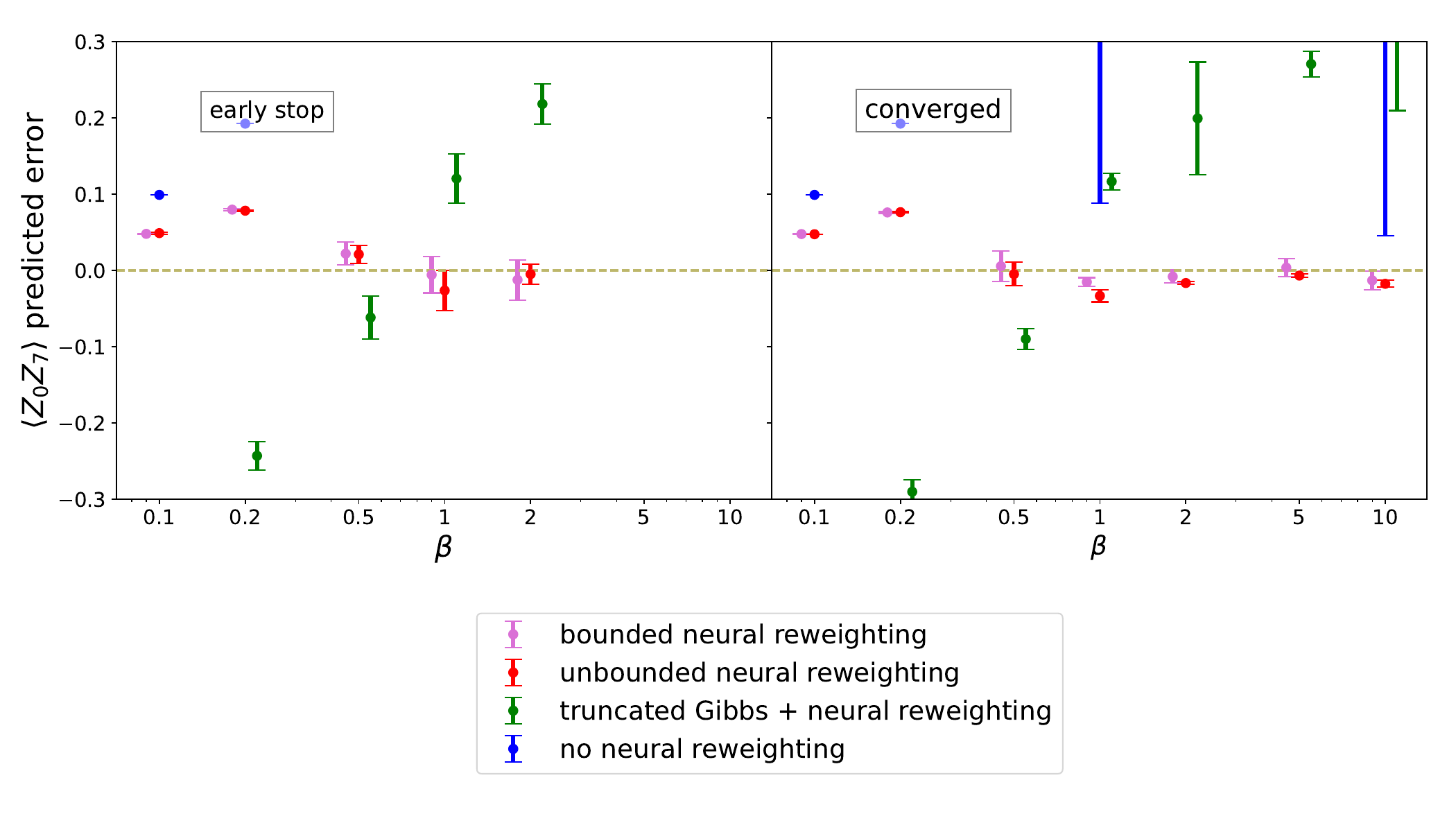}
    \caption{Correlation $\langle Z_0Z_7 \rangle$ prediction error away from exact correlation values. Early stop results are recorded after 300 rounds of gradient descent optimization while converged results are after 2600 rounds of gradient descent optimization. Points beyond the scope of this figure are not shown due to large prediction errors.}
    \label{fig:z0z7}
\end{figure} 

\subsection{Notations}

Quantum gate notation:
\begin{itemize}
\item $R_z$ gate: $e^{i\theta Z}$ where $Z$ is Pauli-Z operator.

\item $R_{zz}$ gate: $e^{i\theta Z_iZ_j}$ where $Z_i$ ($Z_j$) is the Pauli-Z operator on qubit i (j).

\item SWAP gate: the matrix representation is $\text{SWAP} = \begin{pmatrix}1&0&0&0\\0&0&1&0\\0&1&0&0\\0&0&0&1 \end{pmatrix}$.

\item $R_{swap}$ gate: $e^{i\theta \text{SWAP}}$, where $\text{SWAP} = \begin{pmatrix}1&0&0&0\\0&0&1&0\\0&1&0&0\\0&0&0&1 \end{pmatrix}$.
\end{itemize}

    \subsection{Hyperparameter settings}
    
    The neural reweighting probability $f(a)$ is normalized by a softmax activation before the final output as $f(a)=\frac{e^{g(a)}}{\sum_{b} e^{g(b)}}$, where $g(a)$ is a fully connected neural network with output range restricted in $[-e, e]$ for bounded neural reweighting case and $b$ is sum over all bitstring in ancilla qubits. The optimization ends when it occurs $100$ times that the difference of objective function between two optimization rounds is smaller than $10^{-7}$. For other common hyperparameters, see Table.~\ref{tab:hyper}.
    
\begin{table}[t]
    \caption{Hyperparameters in thermal state preparation example.}
    \begin{tabular}{@{}cc@{}}
        \toprule
        Hyperparameters & Value      \\ \midrule
        Optimizers    &  Adam \\ 
        Learning rate for circuit parameters & $0.08*0.5^{s/1200}$  \\
        Learning rate for neural weights & 0.015\\
        Initialization for circuit parameters & N(0, 0.02)\\
        Initialization for neural weights & N(0, 0.005)\\
        \bottomrule
    \end{tabular}
    \label{tab:hyper}
\end{table}

\end{widetext}
	
\end{document}